\documentclass[11pt]{article}
\usepackage{graphicx,hyperref}
\usepackage{amssymb,amsmath,amsthm}
\usepackage{fullpage}
\usepackage[left]{lineno}
\usepackage[]{hyperref}
\newcommand{\comment}[1]{} 

\theoremstyle{definition}

\usepackage{microtype} 
\usepackage{color}

\newcommand\blfootnote[1]{%
  \begingroup
  \renewcommand\thefootnote{}\footnote{#1}%
  \addtocounter{footnote}{-1}%
  \endgroup
}

\title{Maths, Computation and Flamenco: overview and challenges}

\author{J.M. D\'iaz-B\'a\~nez \thanks{Department of Applied Mathematics II, University of Seville, Camino de los descubrimientos s/n, 41092 Seville, Spain. Email: dbanez@us.es}
\and Nadine Kroher
\thanks{Department of Applied Mathematics II, University of Seville, Camino de los descubrimientos s/n, 41092 Seville, Spain. Email: nadine@mxxmusic.com}
}

\begin{document}

\maketitle

%
%
\begin{abstract}
Flamenco is a rich performance-oriented art music genre from Southern Spain which attracts a growing community of aficionados around the globe. Due to its improvisational and expressive nature, its unique musical characteristics, and the fact that the genre is largely undocumented, flamenco poses a number of interesting mathematical and computational challenges. 
Most existing approaches in Musical Information Retrieval (MIR) were developed in the context of popular or classical music and do often not generalize well to non-Western music traditions, in particular when the underlying music theoretical assumptions do not hold for these genres.
Over the recent decade, a number of computational problems related to the automatic analysis of flamenco music have been defined and several methods addressing a variety of musical aspects have been proposed. 
This paper provides an overview of the challenges which arise in the context of computational analysis of flamenco music and outlines an overview of existing approaches.

\vspace{.5cm}

\blfootnote{\begin{minipage}[l]{0.3\columnwidth} \vspace{-3mm}\hspace{-0.41cm} \includegraphics[trim=10cm 6cm 10cm 5cm,clip,scale=0.15]{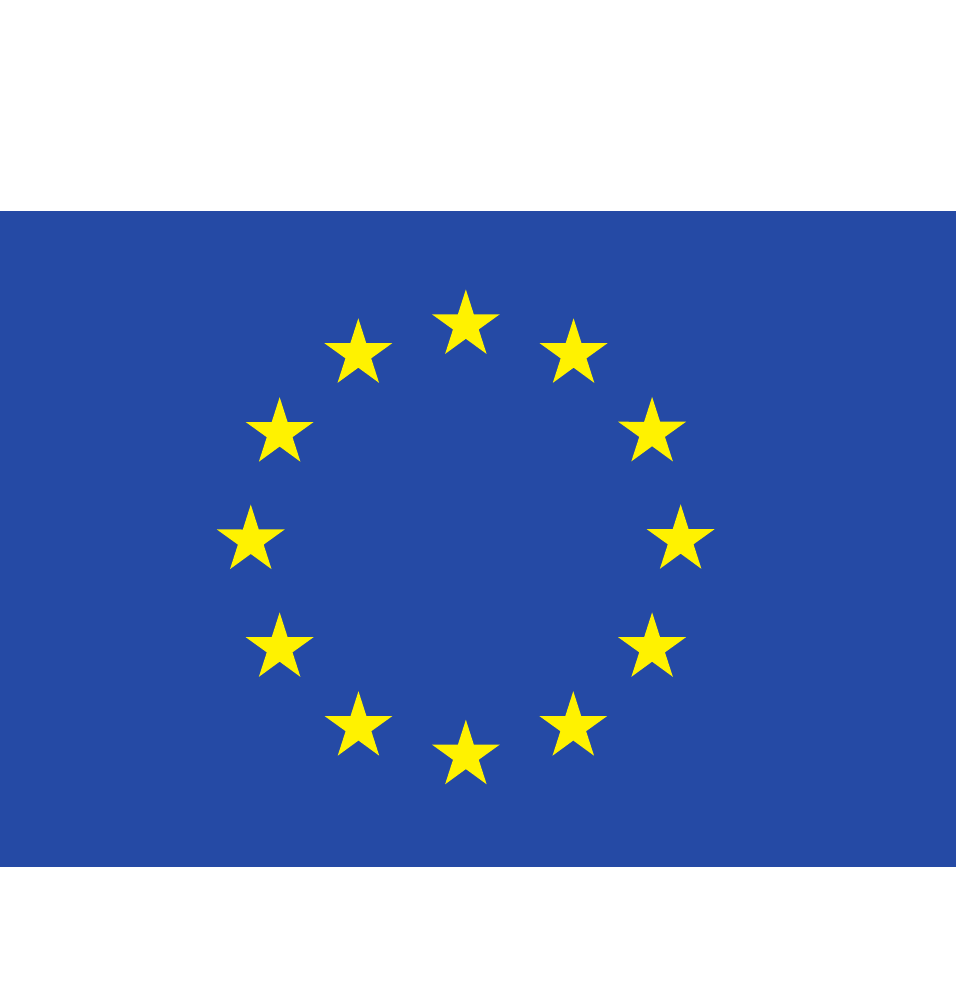} \end{minipage}
\hspace{-3.cm} \begin{minipage}[l][1cm]{0.88\columnwidth} 
		\vspace{-1mm} This research has received funding from the Junta de Andaluc\'ia (project P12-TIC-1362), Spanish Ministry of Economy and Competitiveness (project  MTM2016-76272-R AEI/FEDER,UE), and the European Union's Horizon 2020 research and innovation programme under the Marie Sk\l{}odowska-Curie grant agreement No.\ 734922.
\end{minipage}}
\end{abstract}
\section{Introduction}
The relatively young field of computational analysis of flamenco music has been explored in the scope of the COFLA
research project \cite{cofla} which aims at the development of genre-specific Music Information Retrieval (MIR) methods for flamenco music. Due to its improvisational and expressive nature, its unique musical characteristics, and the fact that the genre is largely undocumented, flamenco poses a number of interesting computational challenges. At the same time, its growing popularity, constant evolution and increasing presence in digital media calls for the development of computational tools to manage existing digital content, allow a broad range of users access to the genre and to contribute to its preservation. In addition, flamenco poses no exception to the ongoing partnership between mathematics and music and the genre has recently been the focus of mathematical research. The development of MIR methods for flamenco, as well as music-theoretical concepts themselves, have given rise to new mathematical problems. A comprehensive elaboration on the synergies between mathematics and flamenco, can be found in \cite{mathAndFlamenco} and a chapter in \cite{montiel} provides a brief overview of flamenco music and its computational study in the context of mathematics. In each section of this paper, we introduce a set of research topics and the arising challenges, describe existing approaches and outlines open problems.

\section{Prerequisites: Data Sets and Automatic Transcription}
A fundamental necessity for computational studies is the creation of data corpora, in our case collections of music recordings, which are not only the basis for experimental evaluation, but are furthermore a valuable resource for data-driven exploratory studies. In the context of flamenco music, there are two well-established and available datasets. The \textit{TONAS}\footnote{\url{https://www.upf.edu/web/mtg/tonas}} dataset contains $72$ recordings of songs belonging to the \textit{ton{\'a}s} style family, together with their respective automatic and manual note-level transcriptions, and annotated style affinity. The \textit{corpusCOFLA} \cite{corpusCOFLA}, which has been developed specifically for computational studies, contains of over 1500 commercial flamenco recordings with manually curated meta-data and several manually annotated subsets. In addition to the audio recordings, the corpus contains editorial and manually curated meta-data, as well as several manually annotated subsets. In addition to audio collections, the knowledge base \textit{FlaBase} \cite{FlaBase}, is a valuable resource for semantic data, including biographical and music-theoretical knowledge gathered from online resources.

The study of various aspects of flamenco singing requires the extraction of the singing voice melody from the audio signal. A comparison of several melody representations in the context of melody classification \cite{CMJPaper} has shown that note-level transcriptions have several advantages over other methods such as the fundamental frequency contour. Since manual note-level transcriptions are highly time consuming, their automatic extraction is an important research topic. A first approach towards computer-assisted transcription of a cappella flamenco recordings was proposed in \cite{SMSTools} and later extended to recordings containing guitar accompaniment \cite{polyTranscription}. Recently, the \textit{CANTE} algorithm, a novel method for the transcription of flamenco singing from monophonic and polyphonic recordings \cite{CANTE}, has shown to yield an improvement compared to \cite{SMSTools}. 

\subsubsection{Open problems:} While automatic transcriptions have been successfully employed in a number of related MIR tasks, the problem can by no means be considered solved. Possible future directions of research include the crowd-sourced creation of a larger annotated dataset, which can then be exploited in the context of machine learning based approaches.

\section{Melodic analysis: retrieval, classification and melodic template extraction}\label{sec:melodyClassification}
A frequently addressed aspect of flamenco music is the existence of melodic templates which set the basis for improvisational performances. In particular, a number of styles and sub-styles are characterized by a specific melodic movement which undergoes heavy ornamentation and variation during individual interpretations. It worth mentioning that this template remains implicit and does not exist in form of a musical score. From a computational perspective, we identify three challenges related to this music theoretical concept: Melody \emph{classification}, \emph{retrieval} and \emph{template extraction}. 

\emph{Melody classification} aims at recognizing the template in a given performance which in turn allows to automatically identify the style or sub-style affinity. Based on the assumption, that performances with a common melodic template are similar to each other and dissimilar to performances based on other templates, the task has mainly been formulated as a \textit{k-nearest-neighbour} classification task, where an unknown performance is labelled based on the labels of its most similar items in an annotated database. In this context, two components are crucial: the way the melody is represented and the employed distance metric.
Early approaches have explored standard distance metrics between manually extracted global mid-level features extracted \cite{melodyClassification10} \cite{melodyClassification16}. While these studies provided valuable insights into the criteria used by flamenco experts to distinguish styles, the need for manual annotations poses a major limitation. Studies targeting fully-automatic systems \cite{melodyClassification14}, \cite{melodyClassification15}, explored \textit{dynamic time warping} alignment between pitch sequences. Acknowledging the quadratic computational cost of this operation both studies investigated several contour simplification algorithms to reduce the sequence length. A systematic comparison of melody representations was conducted in \cite{CMJPaper}, where the advantages of note-level transcriptions were demonstrated. The work furthermore explores unsupervised melody categorization and proposes a set of evaluation strategies, which provide deeper insight into system performance and scalability.

\emph{Melody retrieval} refers to the task of automatically locating a given melodic sequence among a large number of candidates in a digital music collection. Such systems allow users to locate specific items based on their melodic content and furthermore provide the means to explore the evolution of melodic content from an ethno-musicological perspective. From a technical perspective, the retrieval problem focuses on the detection of sub-sections of melodies which are similar to a user query. To this end, \cite{ismir2012} employed a modification of the \textit{context-dependent dynamic time warping algorithm} on the fundamental frequency contour. Since the high amount of micro-tonal detail contained in this representation can distort similarity scores, \cite{fma2016} designed a gap-tolerant alignment method which operates on note-level transcriptions. 

\emph{Melodic template extraction} aims at approximating a the implicit underlying melodic template based on a set of interpretations of the same melody. In \cite{tube}, the task is formulated as a geometric optimization problem where melodies are modelled as polygonal curves in the time-pitch space. the goal is to compute a new polygonal curve, representative of the template, which fits a fixed number $p$ of similar items at each point in time. The particular challenge of this task is that the parameter $p$ corresponds to an amount of melodies, but does not refer to a specific set of melodies. A second approach to the same computational task was proposed in \cite{template} where a progressive multiple sequence alignment procedure is used to construct a graph model holding information on the frequency of notes and note transitions across performances. This model allows the extraction a melodic sequence which approximates the melodic template and allows the computation of a metric for melodic stability. 

\subsubsection{Open problems:} Given the limited amount of ground truth data, the creation of a large annotated audio collection is necessary step for the development of large, scalable melody analysis systems. This requires a significant amount of musicological effort, since a clear style taxonomy, in particular with respect to melodic templates, has so far not been established. 

\section{Detection of phrase-level repetition}\label{sec:repetition}
The structure of European folk music is heavily based on phrase-level repetition. The same observation can be made for certain flamenco styles, which are close to their folkloric origin. Consequently, detecting repeated sung phrases allows us to describe the structure of a performance and can furthermore aid related MIR tasks, such as query by humming and melody compression.
 
 In the context of accompanied flamenco singing, a preliminary audio-based attempt to discover phrase-level repetition was proposed in \cite{eusipco2015}: Sung phrases are detected using a vocal detection algorithm. At a second stage, pair-wise alignment of chroma-based representations of the detected vocal segments is performed and groups of similar phrases were formed using a frame-centric clustering scheme. A more recent work \cite{repeated}, follows the following approach: Automatic transcriptions are segmented into phrases by exploiting certain musical properties typically encountered in folk songs. Then, all pair-wise melodic distances among the detected subsequences are computed, resulting in a distance matrix.  Finally, a standard clustering algorithm receives the computed distance matrix and groups the phrases. As a result, each cluster corresponds to a prototypical melodic pattern and the members of the cluster to the occurrences of the pattern. This method yields convincing results and outperforms the approach in \cite{eusipco2015}.

\subsubsection{Open problems:} The musicological assumptions for the phrase segmentation algorithm of the current system do not generalize to all flamenco styles. In particular, many styles exhibit sung phrases which significantly vary in length. Consequently, a musicological study of phrase boundary characteristics in flamenco music and the development of a genre-tailored algorithm are logical directions for future work. 

\section{Rhythm and its mathematical properties}
Several attempts have been made to computationally model rhythm in flamenco music. In a first approach \cite{diaz2004}, a rhythmic pattern is represented as binary sequences mapped to a circular lattice. 
Based on this representation, mathematical measures can be used to characterize rhythmic similarity between commonly occurring rhythmic patterns and to construct phylogenetic trees to visualize their relationship \cite{diaz2004,gaceta}, which in turn allow us to infer hypothetical ancestral rhythms \cite{caraballo2015finding}. The latter study furthermore generalizes necessary conditions fulfilled by unknown nodes which are useful for computing the ancestral nodes. In \cite{flamencoRhythm09}, different rhythmic similarity measures are evaluated against human judgements.

Mathematical models furthermore allow us to study rhythmic properties and their ethno-musicological meaning. One such property, the rhythmic oddity \cite{chemillier2003}, describes patterns which do not have two beats that divide their span into two intervals of equal duration which implies a notion of asymmetry. In circular notation, this means that no two beats lie diametrically opposite each other on the lattice circle. That is, no two vertices of the polygon are antipodal vertices. The buler\'ia pattern, which is referred to as an asymmetric rhythm in \cite{barba2016}, as well as an antipodal rhythm in \cite{antipodal}, falls into this category.

\subsubsection{Open problems:} While rhythm in flamenco has been analyzed with the used of mathematical models, the automatic detection of beats, downbeats and rhythmic patterns in flamenco recordings remains an open problem. In particular rhythmic pattern detection could significantly improve automatic style detection, since some styles are characterized by specific patterns.

\section{Deep learning based flamenco analysis}
In the context of audio and music processing, deep learning based methods have shown to give promising results, and, in many cases, have been able to outperform state-of-the-art methods. In recent years, several MIR tasks have been approached from a deep learning perspective, including onset detection \cite{deepOnset}, instrument classification \cite{instrRec} and music recommendation \cite{recommender}. To the best of our knowledge, the use of deep learning in the context of automatic content-based description and discovery of flamenco music has been first addressed in \cite{Nadinethesis}, where the focus is on a particular deep learning architecture, the Convolutional Neural Networks (CNNs) and their application to two flamenco-related tasks, \emph{singer identification} in flamenco videos and \emph{structural segmentation} of flamenco music recordings. 

The proposed image-based singer identification system relies of a number of state-of-the-art image processing and computer vision technologies, which are readily available in open source libraries. However, their application to the task at hand goes beyond a trivial re-use of existing techniques, since problem-specific adaptations were necessary and the integration of several processing blocks required a certain level of domain-specific knowledge. The core of the framework is a pre-trained CNN which extracts an embedding with high discriminative power among faces from a given input image. This network was developed and trained for the particular purpose of being re-used in face recognition and authentication tasks, without the need to re-train on a problem-specific candidate set. 

For the second task, a CNN-based system is proposed, which segments a flamenco recording into sections of consistent instrumentation. The method and has shown to outperform a baseline method which uses an ensemble of shallow classifiers. Moving towards the area of data mining, the segmentation backend was applied to the analysis of a large corpus of commercial flamenco recordings, exploring the resulting automatic annotations in a data-driven study. The goal is to enable musicological studies that would otherwise require time-consuming manual procedures, and verify, at a large scale, via computational means and a data-driven approach, existing musicological observations. This computational study, which revealed a number of trends and correlations, is the first of its kind in the context of flamenco music.

\subsubsection{Open problems:} The work in \cite{Nadinethesis} has demonstrated the potential of deep learning for the analysis of flamenco. Provided that sufficient amounts of annotated data exist, this methodology could be applied to other tasks in the context of flamenco, including beat tracking and automatic transcription.  

\section{Conclusions}\label{conclusions}
We have outlined the ongoing research efforts towards the development genre-specific MIR methods for flamenco music. For each of the commonly addressed task, we have presented existing work and outlined current shortcomings, open problems and future extensions. Beyond the tasks described above, several domains of flamenco music have so far not been explored from an computational or mathematical viewpoint. Possible new lines of research could for example include the analysis of flamenco dance through computer vision or sensors. Furthermore, past approaches have mainly focused on the singing voice and less attention has been paid to the guitar accompaniment.

\end{document}